\documentclass[prb,twocolumn]{revtex4-1} 

\usepackage{amsmath}  
\usepackage{amsfonts} 
\usepackage{graphicx} 
\usepackage{siunitx} 
\usepackage{verbatim}
\usepackage{color}

\begin{document}


\title{Modeling and measuring the non-ideal characteristics of transmission lines}


\author{J. S. Bobowski}
\email{jake.bobowski@ubc.ca} 
\affiliation{Department of Physics, University of British Columbia, Kelowna, British Columbia, Canada V1V 1V7}


\date{\today}

\begin{abstract}
We describe a simple method to experimentally determine the frequency dependencies of the per-unit-length resistance and conductance of transmission lines.  The experiment is intended as a supplement to the classic measurement of the transient response of a transmission line to a voltage step or pulse.  In the transient experiment, an ideal (lossless) model of the transmission line is used to determine the characteristic impedance and signal propagation speed.  In our experiment, the insertion losses of various coaxial cables are measured as a function of frequency from \SI{1}{} to \SI{2000}{\mega\hertz}.  A full distributed circuit model of the transmission line that includes both conductor and dielectric losses is needed to fit the frequency dependence of the measured insertion losses.  Our model assumes physically-sensible frequency dependencies for the per-unit-length resistance and conductance that are determined by the geometry of the coaxial transmission lines used in the measurements.      
\end{abstract}

\maketitle 

\section{Introduction}

Lumped-element circuit analysis fails when the wavelengths of the signals of interest approach the size of the circuit elements and/or connecting wires.  In this limit, the voltage and current along, for example, the length of a pair of wires are not uniform and a distributed circuit model of the wires must be used to properly analyze the circuit behavior.\cite{Haus:1989, Pozar:2012, Collier:2013, Snoke:2015}  These so-called transmission line effects are rich in physics and, in many cases, can defy common intuition.  Furthermore, due to the ever decreasing size of circuits and increasing data rates, transmission line effects are more prevalent than ever.\cite{Paul:2009} 

From a pedagogical standpoint, transmission lines are commonly used when deriving the expression for the thermal noise radiated by a resistor.\cite{Nyquist:1928}  In the derivation a transmission line, with both ends terminated by matched load resistors, is treated as a 1-D blackbody.  The thermal power radiated by one resistor is completely absorbed by the other and the emitted radiation satisfies standing wave conditions (normal modes) set by the length of the transmission line.  Transmission lines models have also been used to analyze problems in thermodynamics and mechanics.  An $RC$ transmission line circuit model has been used to understand diffusion of heat along the length of a conducting bar,\cite{Greenslade:1989} and an analogy has been made between a system of coupled pendula and coupled transmission lines.\cite{Teoh:1996}  

It is also worth pointing out that there have been very clever uses of discrete transmission lines.  In one example, reverse-biased variable capacitance diodes (varactor diodes) were used to construct a nonlinear transmission line that supports solitons.\cite{Kuusela:1987}  In a second example, the capacitance in some sections of the discrete transmission line was changed to mimic a change in refractive index.  These structures were used to experimentally demonstrate the principles behind highly-reflective dielectric mirrors and confinement of electromagnetic (EM) waves by Bragg reflection.\cite{Janssen:1988}  

Many of the undergraduate transmission line experiments described in the literature focus on the propagation of short pulses along the length of a line that is assumed to be lossless.  These measurements allow students to determine the propagation speed of the pulse and observe phase changes resulting from reflections at various load terminations.  By tuning the load resistance to eliminate the reflections, students can also estimate the characteristic impedance of the transmission line.\cite{Rank:1969, Holuj:1982, Watson:1995, Mak:2003}  Another experiment in which dissipative effects are typically neglected is the transient response of a transmission line to an applied pulse that is many times longer than the time required to travel the length of the line.\cite{Collier:2013, Haus:1989}  This measurement is particularly interesting because a wide variety of intricate voltage transients are possible depending on the values of the source and load impedances used.  In this paper, we demonstrate one of the simpler cases using an open-circuit load and a large source resistance.

\begin{figure*}
\centering{
\begin{tabular}{c}
(a)\includegraphics[keepaspectratio, width=2\columnwidth]{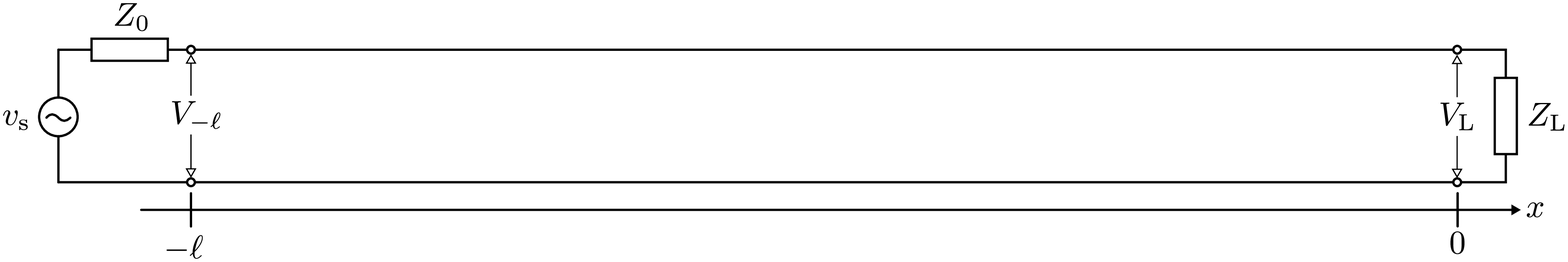}\\
~ \\
(b)\includegraphics[keepaspectratio, width=2\columnwidth]{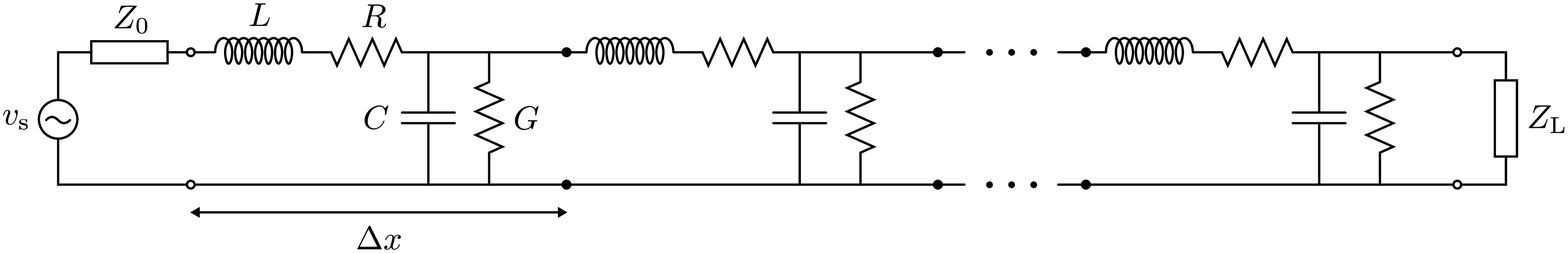}
\end{tabular}
}
\caption{\label{fig:distributed}(a) A transmission line of length $\ell$ connected to a signal source with output impedance $Z_0$ at $x=-\ell$ and a load impedance $Z_\mathrm{L}$ at $x=0$. (b) The distributed circuit model of a transmission line.}
\end{figure*}

After using the measured transient response to determine the characteristics of a lossless line, we turn to our main focus which is to quantitatively measure and analyze transmission line dissipation due to conductor and dielectric losses.  There are examples of undergraduate laboratories in which the dissipation due to a length of transmission line has been measured.\cite{Colpitts:2002, Serra:2004}  However, in these cases, after making an attenuation measurement, there is little to no discussion about the origins of the dissipation and the relative importance of the various sources of loss.  One of the objectives of this paper is to make a simple, yet reasonably precise, measurement of the power dissipation due to a length of transmission line over a wide frequency range.  Another, more significant, objective is to use the data and physical insights to quantitatively determine the relative magnitudes of conductor and dielectric losses as a function of frequency. 

The pedagogical benefits of our work are as follows: (1) We describe two experiments that yield precise results, yet are relatively simple and can be done within the usual three-hour period of a university teaching laboratory. (2) Some of the theory needed to interpret the experimental results is typically not explicitly covered in an undergraduate physics program.  Therefore, students are required to research and develop some of the relevant theory themselves. (3) The main focus is characterizing dissipation in transmission lines which is of great practical importance, but not often emphasized in a classroom setting. (4) In order to fit the attenuation data, students must develop models for conductor and dielectric losses in a coaxial cable.  These exercises require students to activate prior knowledge which gets integrated it into their models and leads to new physical insights.

The outline of the paper is as follows: In Sec.~\ref{sec:model} the distributed circuit model of a transmission line and some of its important features are reviewed.  Section~\ref{sec:transient} presents the transient response of transmission lines to a long-duration pulse.  The experimental results are used to determine the per-unit-length capacitance and inductance of ideal (lossless) transmission lines.  Dissipative effects are treated in Sec.~\ref{sec:dissipation}.  First, the expected frequency response of the transmission line insertion loss is calculated and then the experimental measurements are presented.  In Sec.~\ref{sec:losses}, physically-motivated models for the frequency dependencies of the per-unit-length resistance and conductance of coaxial transmission lines are developed.  These models are then used to fit the experimentally-measured insertion loss in Sec.~\ref{sec:fitting}.  Finally, the main results are summarized in Sec.~\ref{sec:summary}.  

\section{Distributed Circuit Model of a Transmission Line}\label{sec:model}
Figure~\ref{fig:distributed}(a) shows a transmission line of length $\ell$.  One end, at $x=-\ell$, is connected to a signal source that has output impedance $Z_0$ and the opposite end, at $x=0$, is terminated by load impedance $Z_\mathrm{L}$.  The equivalent distributed circuit model of the transmission line is shown in Fig.~\ref{fig:distributed}(b).  It is made up of $n=\ell/\Delta x$ daisy-chained segments, where $\Delta x$ is the length of each segment.  The segments have series inductance and resistance $L\Delta x$ and $R\Delta x$, respectively and shunt capacitance and conductance $C\Delta x$ and $G\Delta x$, respectively.  In this model, $R$ represents conductor losses and $G$ accounts for dielectric losses.  The distributed circuit model describes the behavior of real transmission lines in the limit that the number of segments $n\to\infty$ or, equivalently, $\Delta x\to 0$.

We now state some of the important results arising from an analysis of the distributed circuit model.  Our analysis is presented in terms of Laplace transforms which allow one to deduce both the transient and steady-state responses of transmission lines.~\cite{Shenkman:2005, Griffith:1990}   We also refer the reader to alternative treatments given in Refs.~\onlinecite{Haus:1989, Pozar:2012, Collier:2013} with additional insights provided in Ref.~\onlinecite{Roy:2010}.  In the limit $\Delta x\to 0$, an analysis of the Kirchhoff loop and junction rules for a single segment of the transmission line circuit model leads to a pair of coupled first-order partial differential equations in voltage and current known as the telegrapher's equations
\begin{align}   
\frac{\partial v}{\partial x}&=-L\frac{\partial i}{\partial t}-Ri\\
\frac{\partial i}{\partial x}&=-C\frac{\partial v}{\partial t}-Gv,
\end{align}
where $v=v\left(x,t\right)$ and $i=i\left(x,t\right)$.  Taking the Laplace transform, defined as \mbox{$F\left(x,s\right)=\int_o^\infty f(x,t)e^{-st}dt$}, leads to the following representation of the of the telegrapher's equations in the complex frequency, or $s$, domain
\begin{align}
\frac{\partial V}{\partial x}&=-\left(R+sL\right)I\label{eq:LT1}\\
\frac{\partial I}{\partial x}&=-\left(G+sC\right)V,\label{eq:LT2}
\end{align}
where, for simplicity, we have assumed that the initial transmission line voltage and current are everywhere zero.  Equations (\ref{eq:LT1}) and (\ref{eq:LT2}) can be re-expressed as a pair of independent second-order partial differential equations
\begin{align}
\frac{\partial^2 V}{\partial x^2} &=\gamma^2 V\label{eq:d2Vdx2}\\
\frac{\partial^2 I}{\partial x^2} &=\gamma^2 I,\label{eq:d2Idx2}
\end{align} 
where $\gamma \equiv \sqrt{\left(R+sL\right)\left(G+sC\right)}$.  The solutions to Eqs.~(\ref{eq:d2Vdx2}) and (\ref{eq:d2Idx2}) are traveling waves
\begin{align}
V(x,s) &=V_+e^{-\gamma x}+V_-e^{\gamma x}\label{eq:Vx}\\
I(x,s) &=\frac{1}{Z_\mathrm{c}}\left[V_+e^{-\gamma x}-V_-e^{\gamma x}\right],\label{eq:Ix}
\end{align}
where $V_+$ and $V_-$ terms represent signals propagating in the $+x$ and $-x$ directions, respectively, and
\begin{equation}
Z_\mathrm{c}=\sqrt{\frac{R+s L}{G+s C}},\label{eq:Zc}
\end{equation}
is the characteristic impedance of the transmission line.  The $V_+$ and $V_-$ coefficients are related via $V_-=\Gamma V_+$ where
\begin{equation}
\Gamma=\frac{Z_\mathrm{L} - Z_\mathrm{c}}{Z_\mathrm{L} + Z_\mathrm{c}},\label{eq:G}
\end{equation} 
is the reflection coefficient determined by the mismatch between $Z_\mathrm{c}$ and the load termination.  Equation~(\ref{eq:G}) is obtained by requiring $V(0)/I(0)=Z_\mathrm{L}$ at the load termination.

The steady state solutions for harmonic voltages and currents of angular frequency $\omega$ are given by Eqs.~(\ref{eq:Vx}) and (\ref{eq:Ix}) with $s\to j\omega$.  In this case, $V$ and $I$ represent the voltage and current amplitudes along the length of the transmission line and the propagation constant \mbox{$\gamma=\alpha+j\beta$} is complex, where $\alpha$ and $\beta$ are the attenuation and phase constants, respectively.  The time-dependent transient solutions, on the other hand, can be obtained by taking the inverse Laplace transform of $V\left(x,s\right)$ and $I\left(x,s\right)$.\cite{Shenkman:2005, Griffith:1990}

\subsection{Lossloss transmission lines}
In a lossless transmission line $R$ and $G$ are assumed to be negligible.  In this limit, the propagation constant and characteristic impedance become
\begin{align}
\gamma &=s\sqrt{LC}=s/v_0\\
Z_\mathrm{c} &=\sqrt{L/C},
\end{align}
where $v_0=1/\sqrt{LC}$ is the signal propagation speed.  Note that, if $v_0$ and $Z_\mathrm{c}$ are measured, then the inductance per unit length and capacitance per unit length of the transmission line can be determined via \mbox{$L=Z_\mathrm{c}/v_0$} and \mbox{$C=\left(v_0Z_\mathrm{c}\right)^{-1}$}.

\section{Transient Response}\label{sec:transient}
\begin{figure}
\centering{
\begin{tabular}{c}
(a)~\includegraphics[keepaspectratio, width=0.92\columnwidth]{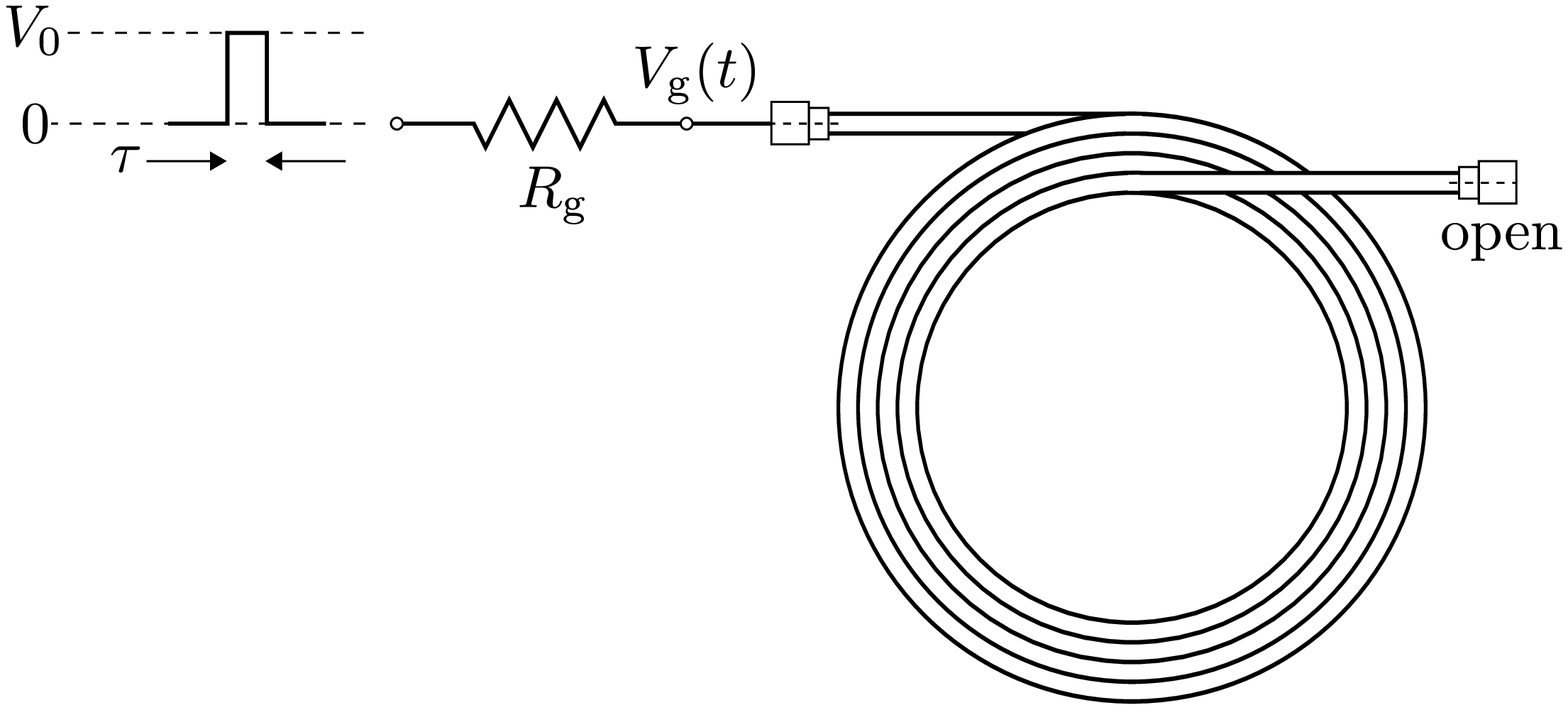}\\
~ \\
(b)~\includegraphics[keepaspectratio, width=0.92\columnwidth]{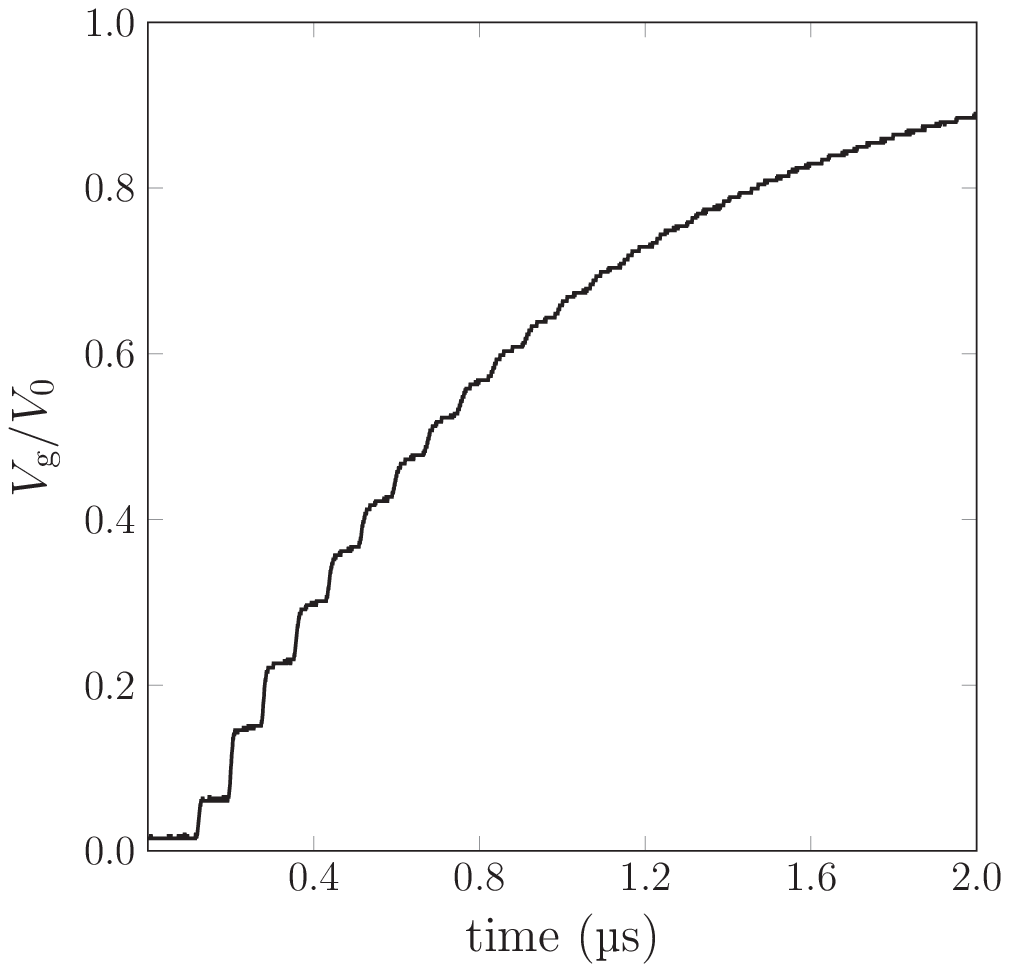}
\end{tabular}
}
\caption{\label{fig:transients}(a) Schematic diagram of the experimental setup used to measure the transient response of a transmission line.  
(b) The transient response of a long semi-rigid UT-141 coaxial cable.  The voltage $V_\mathrm{g}(t)$ has been scaled by the maximum voltage reached a long time after the pulse is applied.  
Although data was collected up to \SI{10}{\micro\second}, only the data up to \SI{2}{\micro\second} are shown in order to highlight the step-like features in $V_\mathrm{g}(t)$.}
\end{figure}

We now consider a length of lossless transmission line with one end open ($Z_\mathrm{L}\to\infty$) and the opposite end connected to a resistance $R_\mathrm{g}\gg Z_\mathrm{c}$.  A square pulse of height $V_0$ is applied to the free end of $R_\mathrm{g}$. The width of pulse is chosen to be very long compared to the time $\ell/v_0$ that it takes signals to travel the length of the transmission line.  After the pulse is applied, the time evolution of the voltage $V_\mathrm{g}$ at the junction between $R_\mathrm{g}$ and the transmission line is measured.  The experimental setup is shown schematically in Fig.~\ref{fig:transients}(a).  To ensure a well-defined pulse shape, $R_\mathrm{g}$ is chosen to be much larger than the output impedance of the pulse generator.  In our experiments, we used $R_\mathrm{g}=\SI{1}{\kilo\ohm}$ and an HP 8011A pulse generator with a \SI[number-unit-product=\text{-}]{50}{\ohm} output impedance.  A list of the equipment needed to perform all of the measurements described in this paper along with possible vendors and cost estimates is given in the appendix.

\begin{figure*}
\centering{
\begin{tabular}{cc}
(a)\includegraphics[width=0.9\columnwidth]{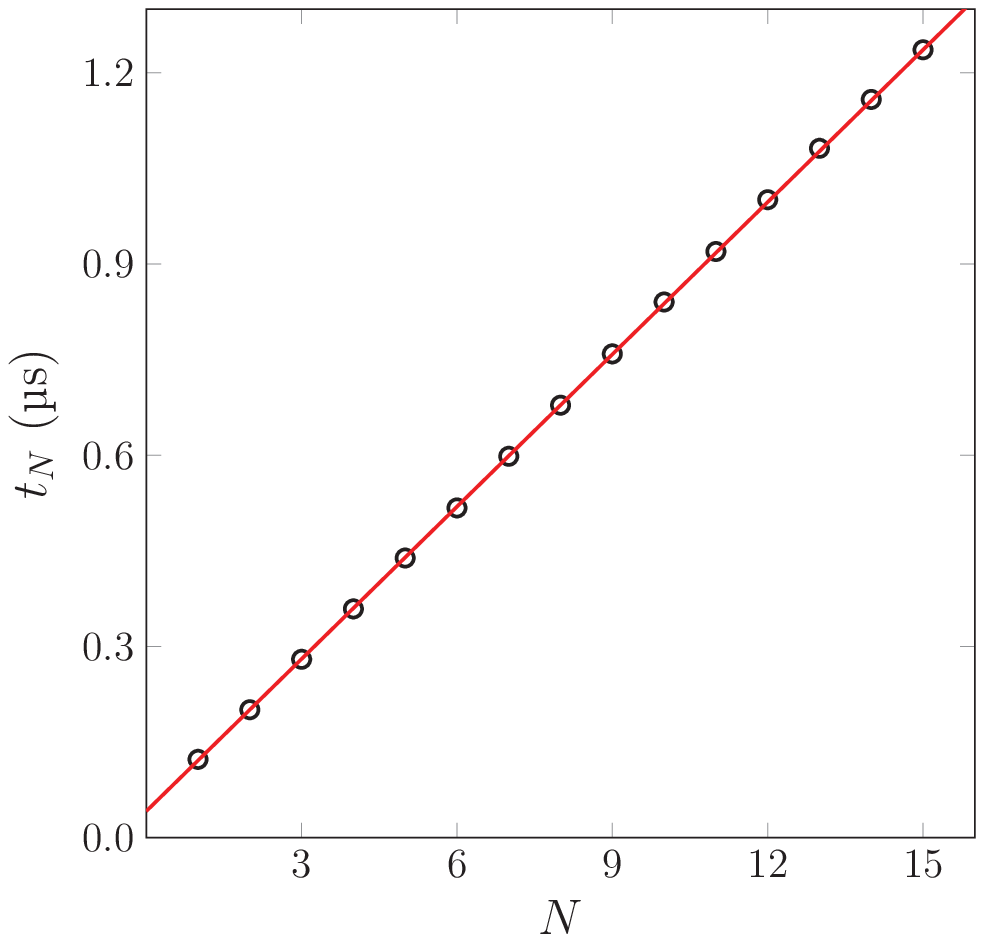} & \qquad(b)\includegraphics[width=0.96\columnwidth]{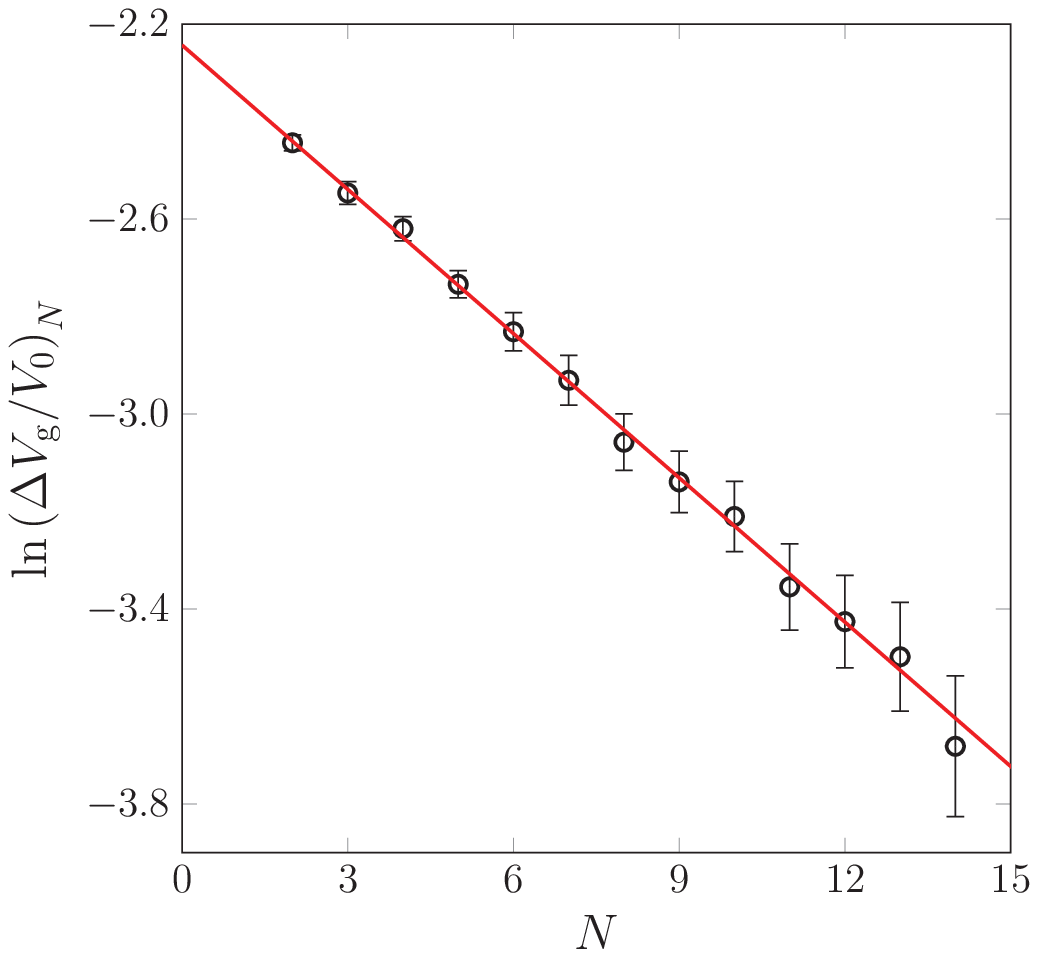}
\end{tabular}
}
\caption{\label{fig:transFits}Analysis of the transient response of the semi-rigid coaxial cable. (a) Step time versus step number.  The error bars are smaller than the point size.  The red line is a linear fit to the data.  
(b) Plot of $\ln\left(\Delta V_\mathrm{g}/V_0\right)_N$ as a function of step number.  The red line is a linear fit to the data.}
\end{figure*}

Although we do not provide a full analysis of this problem here, we refer the reader to the treatment given in Sec.~14.4 of Ref.~\onlinecite{Haus:1989}.  Figure~\ref{fig:transients}(b) shows a measurement of $V_\mathrm{g}$ as a function to time using a \SI[number-unit-product=\text{-}]{8.07}{\meter} length of semi-rigid UT-141 coaxial cable.  The data were recorded using a Tektronix TBS 1104 digital oscilloscope (\SI{100}{\mega\hertz} bandwidth).  The step-like pattern observed in $V_\mathrm{g}(t)$ is due to repeated reflections at the two ends of the transmission line.  The reflection coefficient at the open end is $\Gamma=1$ and at the source end it is given by \mbox{$\Gamma_\mathrm{g}=\left(R_\mathrm{g}-Z_\mathrm{c}\right)/\left(R_\mathrm{g}+Z_\mathrm{c}\right)\approx 0.90$}.

The time between adjacent steps is $\Delta t=2\ell/v_0$ which corresponds to the time required for signals to travel twice the length of the line.  A plot of the time of the $N^\mathrm{th}$ step versus $N$ results in a straight line with slope \mbox{$m_1=\Delta t$} which can then be used to determine the signal propagation speed as has been done in Fig.~\ref{fig:transFits}(a).  

For steps $N\ge 2$, the change in $V_\mathrm{g}$ is given by
\begin{equation}
\left(\frac{\Delta V_{\mathrm{g}}}{V_0}\right)_N=\frac{1-\Gamma_\mathrm{g}^2}{2\Gamma_\mathrm{g}}\Gamma_\mathrm{g}^N,
\end{equation}
such that, because $0<\Gamma_\mathrm{g}<1$, the voltage steps decrease in size as $N$ increases.\cite{Haus:1989, Langlois:1995}  A plot of $\ln\left(\Delta V_\mathrm{g}/V_0\right)_N$ versus $N$ is linear with slope $m_2=\ln\Gamma_\mathrm{g}$ which allows for a determination of the characteristic impedance $Z_\mathrm{c}$ of the transmission line.

Figures~\ref{fig:transFits}(a) and (b) show results of the analysis of the transient voltage steps for the semi-rigid coaxial cable.  The slopes of the linear fits and the corresponding values of $v_0$ and $Z_\mathrm{c}$ are given in Table~\ref{tab:trans}.  The table also includes determinations of $C$, $L$, and the dielectric constant \mbox{$\varepsilon^\prime=\left(c/v_0\right)^2$} of the insulator separating the inner and outer conductors of the coaxial cables.  Although not shown, the transient responses of an RG-58 BNC coaxial cable and a high-quality (HQ) sma coaxial cable were also measured.  As will be shown in Sec.~\ref{sec:fitting}, the conductor and dielectric losses of the HQ cable are low compared to those of the other cables measured.  This cable, of unknown make and model, was donated to the undergraduate lab and has \SI[number-unit-product=\text{-}]{9}{\milli\meter} diameter, a cloth-woven jacket, and is fitted with specialty sma connectors.  
\begin{table}\caption{\label{tab:trans}Parameters extracted from an analysis of the transient response of various coaxial cables.}
\begin{tabular}{llll}
\hline\hline\\ [-1.5ex]
~\qquad~~\qquad~~ & {\bf RG-58}\qquad~~~ & {\bf UT-141}\qquad~~~ & {\bf HQ sma}\\[0.5ex]
\hline\\ [-1.5ex]
$\ell$ (\SI{}{\meter}) & $7.61\pm 0.02$ & $8.07\pm 0.03$ & $8.04\pm 0.02$\\
$m_1$ (\SI{}{\nano\second}) & $79.9\pm 0.2$ & $79.6\pm 0.1$ & $69.64\pm 0.08$\\
$v_0/c$ & $0.635\pm 0.002$ & $0.676\pm 0.003$ & $0.770\pm 0.002$\\
$\varepsilon^\prime$ & $2.48\pm 0.02$ & $2.19\pm 0.02$ & $1.69\pm 0.01$\\
$\left\vert m_2\right\vert$ & $0.107\pm 0.002$ & $0.099\pm 0.001$ & $0.100\pm 0.001$\\
$Z_\mathrm{c}$ (\SI{}{\ohm}) & $53.2\pm 0.9$ & $49.3\pm0.6$ & $49.8\pm 0.7$\\
$C$ (\SI[per-mode=repeated-symbol]{}{\pico\farad\per\meter}) & $99\pm2$ & $100\pm 1$ & $87\pm 1$\\
$L$ (\SI[per-mode=repeated-symbol]{}{\nano\henry\per\meter}) & $279\pm 5$ & $243\pm 3$ & $216\pm 3$\\[0.5ex]
\hline\hline
\end{tabular}
\end{table} 

The RG58C/U coaxial cable specifications from Pasternack give a velocity of propagation that is $0.659c$ and a capacitance of $\SI[per-mode=repeated-symbol]{101.05}{\pico\farad\per\meter}$, both of which are in reasonably good agreement with our results.\cite{Pasternack:2017}  On the other hand, Pasternack specifies a characteristic impedance of \SI{50}{\ohm} which is about 3.5 standard deviations away from our measurement.  We will discuss a possible reason for this difference in Sec.~\ref{sec:fitting}.  The specifications for the UT-141-HA-M17 semi-rigid coaxial cable from Micro-Coax are $v_0=0.7c$, $C=\SI[per-mode=repeated-symbol]{98.1}{\pico\farad\per\meter}$, and $Z_\mathrm{c}=\SI[separate-uncertainty = true, multi-part-units=single]{50\pm 1}{\ohm}$, all of which are in reasonable agreement with the results given in Table~\ref{tab:trans}.\cite{MicroCoax}  

Figure~\ref{fig:transients}(b) shows a rounding of the steps that becomes more pronounced as time evolves.  When the transient responses of the three cables are compared, there are no obvious differences in the observed rounding.  This effect may be due to the onset of higher-order modes in the coaxial transmission lines.  The TE$_{11}$ mode has the lowest cutoff frequency given by $f_\mathrm{c}\approx 0.340v_0/(r_2+r_1)$ where $r_1$ is the radius of the center conductor and $r_2$ is the inner radius of the outer conductor.\cite{Collier:2013}  For the semi-rigid coaxial cable, with $r_1=\SI{0.460}{\milli\meter}$ and $r_2=\SI{1.493}{\milli\meter}$, the TE$_{11}$ cutoff frequency is approximately \SI{35}{\giga\hertz}. 

\section{Dissipation and Frequency Response}\label{sec:dissipation}
We now consider the simple scenario depicted in Fig.~\ref{fig:distributed}(a).  The signal source outputs a sinusoidal wave of angular frequency $\omega$ and we wish to calculate the power delivered to the load impedance $Z_\mathrm{L}$.  We assume that both $R$ and $G$ are small but not negligible and consider the steady-state solutions given by Eqs.~(\ref{eq:Vx}) and (\ref{eq:Ix}).

Assuming that the second order cross term $RG$ in $\gamma$ is small and that, at all frequencies of interest, \mbox{$R/\left(\omega L\right)+G/\left(\omega C\right)\ll 1$}, the propagation constant can be approximated as
\begin{equation}
\gamma\approx j\frac{\omega}{v_0}+\frac{1}{2}\left(\frac{R}{Z_0}+GZ_0\right)\equiv j\frac{\omega}{v_0}+\frac{1}{2}\alpha_+,\label{eq:gammaApprox}
\end{equation}
where, as before, $v_0=1/\sqrt{LC}$ and we denote the characteristic impedance of a lossless transmission as \mbox{$Z_0=\sqrt{L/C}$}.  As is usaully the case, the output impedance of the signal generator is assumed to also be equal to $Z_0$.  Using the same approximations, the characteristic impedance of a lossy transmission line given by Eq.~(\ref{eq:Zc}) with $s=j\omega$ can be written as
\begin{align}
Z_\mathrm{c} &\approx Z_0\left[1-\frac{jv_0}{2\omega}\left(\frac{R}{Z_0}-GZ_0\right)\right]\nonumber\\
&\equiv Z_0\left(1-\frac{jv_0}{2\omega}\alpha_-\right).\label{eq:ZcApprox}
\end{align}  
In Eqs.~(\ref{eq:gammaApprox}) and (\ref{eq:ZcApprox}), the quantities \mbox{$\alpha_\pm\equiv\left(R/Z_0\pm GZ_0\right)$} have been defined.  Next, using Eq.~(\ref{eq:Vx}) and the fact that $V_-=\Gamma V_+$, $V_+$ can be expressed in terms of the voltage amplitude at $x=-\ell$: \mbox{$V_+=V_{-\ell}\left[e^{\gamma\ell}+\Gamma e^{-\gamma\ell}\right]^{-1}$}.  Substituting this result for $V_+$ back into Eq.~(\ref{eq:Vx}) and using Eq.~(\ref{eq:G}) for $\Gamma$ allows one to solve for the voltage amplitude $V_\mathrm{L}=V\left(0\right)$ at the load impedance $Z_\mathrm{L}$
\begin{equation}
\frac{V_\mathrm{L}}{V_{-\ell}}=2\left[\left(e^{\gamma\ell}+e^{-\gamma\ell}\right)+\frac{Z_\mathrm{c}}{Z_\mathrm{L}}\left(e^{\gamma\ell}-e^{-\gamma\ell}\right)\right]^{-1}.
\end{equation} 
Substituting in the approximate forms of $\gamma$ and $Z_\mathrm{c}$ from Eqs.~(\ref{eq:gammaApprox}) and (\ref{eq:ZcApprox}) yields
\begin{widetext}
\begin{align}
\left(\frac{V_\mathrm{L}}{V_{-\ell}}\right)^{-1}=\left[\cos\frac{\omega\ell}{v_0}\cosh\frac{\alpha_+\ell}{2}+\frac{Z_0}{Z_\mathrm{L}}\left(\cos\frac{\omega\ell}{v_0}\sinh\frac{\alpha_+\ell}{2}-\frac{v_0\alpha_-}{2\omega}\sin\frac{\omega\ell}{v_0}\cosh\frac{\alpha_+\ell}{2}\right)\right]\nonumber\\
+j\left[\sin\frac{\omega\ell}{v_0}\sinh\frac{\alpha_+\ell}{2}+\frac{Z_0}{Z_\mathrm{L}}\left(\sin\frac{\omega\ell}{v_0}\cosh\frac{\alpha_+\ell}{2}+\frac{v_0\alpha_-}{2\omega}\cos\frac{\omega\ell}{v_0}\sinh\frac{\alpha_+\ell}{2}\right)\right].\label{eq:long}
\end{align}
\end{widetext}
Identifying $p$ and $q$ as the real and imaginary parts of $\left(V_\mathrm{L}/V_{-\ell}\right)^{-1}$, respectively, allows one to calculate \mbox{$\left\vert V_\mathrm{L}/V_{-\ell}\right\vert=\left(p^2+q^2\right)^{-1/2}$} and $\tan\phi=-q/p$, where $\phi$ is phase difference between the voltages at $x=-\ell$ and $x=0$.  In the lossless limit, $\alpha_+=\alpha_-=0$, Eq.~(\ref{eq:long}) reduces to
\begin{equation}
\frac{V_\mathrm{L}}{V_{-\ell}}=\left[\cos\frac{\omega\ell}{v_0}+j\frac{Z_0}{Z_\mathrm{L}}\sin\frac{\omega\ell}{v_0}\right]^{-1},
\end{equation}
such that, as expected, $\left\vert V_\mathrm{L}/V_{-\ell}\right\vert=1$ when $Z_\mathrm{L}=Z_0$.

Our objective was to measure the ratio of the signal power at $x=0$ to the power at $x=-\ell$ as a function of frequency and then compare it to $\left\vert V_\mathrm{L}/V_{-\ell}\right\vert^2$ calculated from Eq.~(\ref{eq:long}).  This comparison requires models for the frequency dependencies of the per-unit-length resistance and conductance of the coaxial transmission lines used in our measurements.  These models are developed in the next section.

\subsection{Models of resistance and conductance}\label{sec:losses}
\begin{figure}
\centering{
\begin{tabular}{c}
(a)\includegraphics[width=0.7\columnwidth]{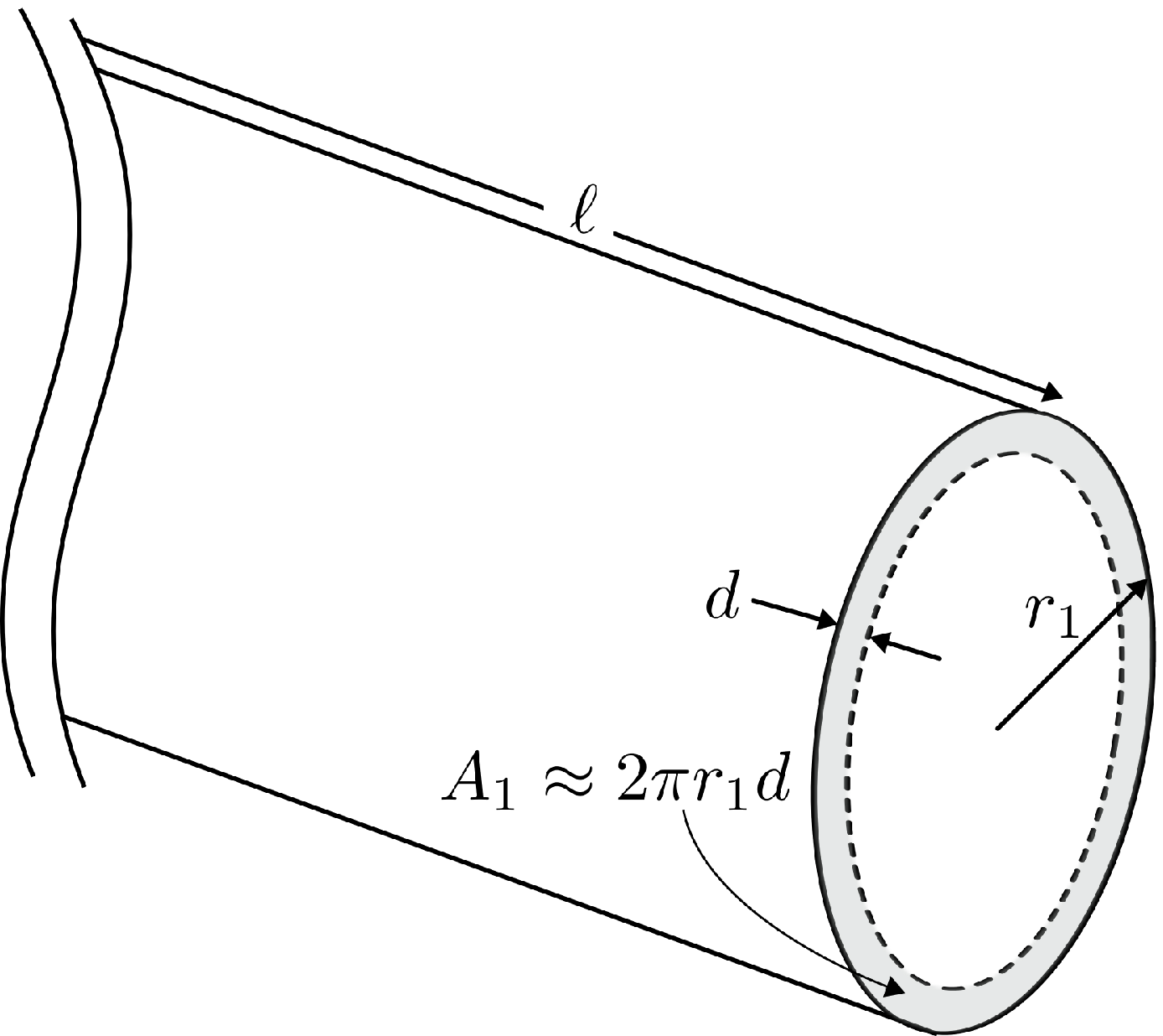}\\
~ \\
(b)\includegraphics[width=0.7\columnwidth]{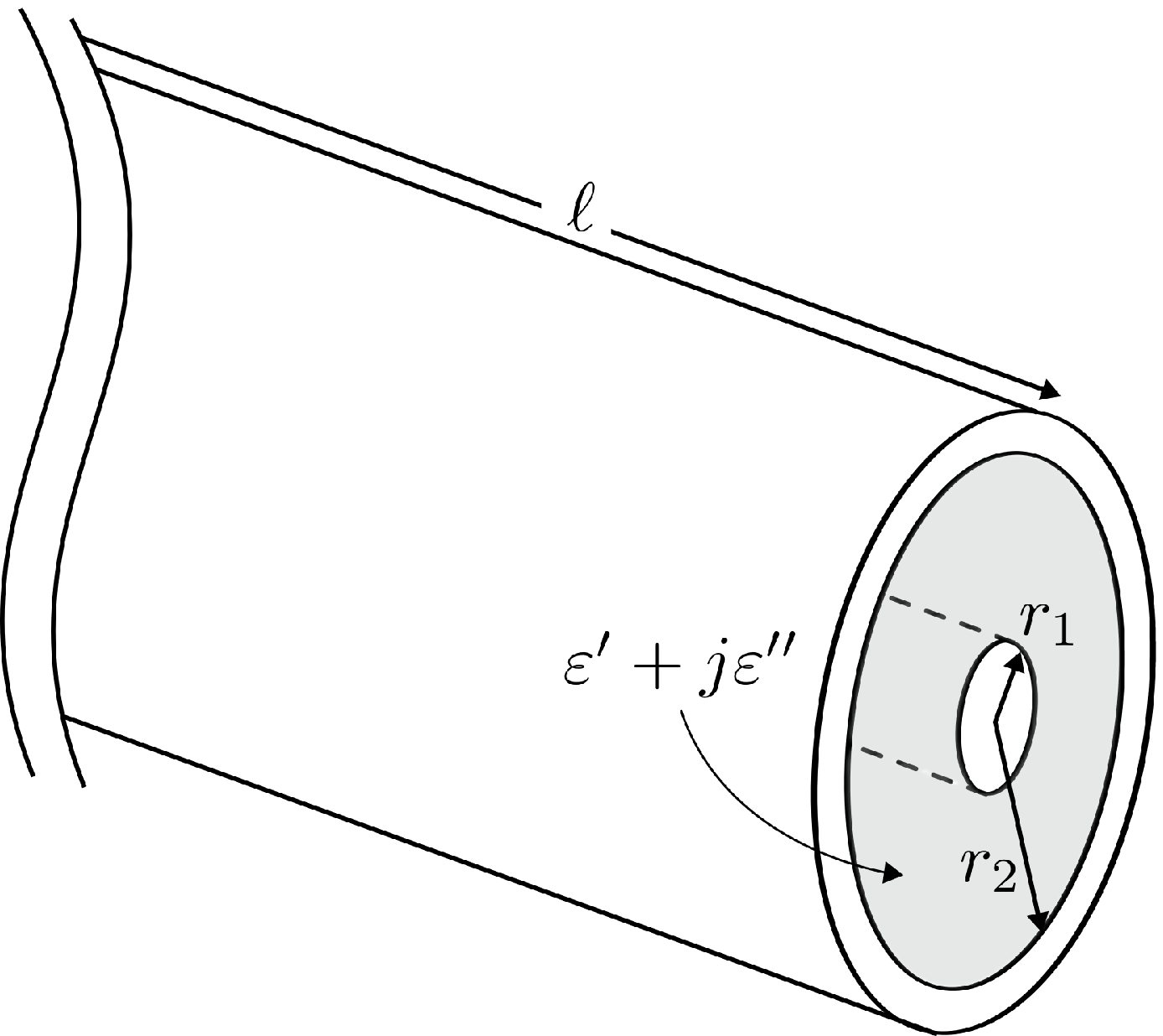}
\end{tabular}
}
\caption{\label{fig:RandG}(a) Schematic drawing of the center conductor of a coaxial cable of radius $r_1$.  The shaded region depicts the cross-sectional area through which the current travels which is determined by the frequency-dependent EM skin depth $d$.  (b) Schematic drawing of a coaxial cable.  The space between the center and outer conductors is filled with a dielectric material (shaded region) with complex relative permittivity \mbox{$\varepsilon_\mathrm{r}=\varepsilon^\prime - j\varepsilon^{\prime\prime}$}.  
}
\end{figure}

First, we consider the resistance which is due to the usual Joule heating in conductors.  Figure~\ref{fig:RandG}(a) shows a schematic diagram of a section of the center conductor from a coaxial cable.  The current in the center conductor is restricted to a region that is within an EM skin depth $d$ of the surface. For a good conductor
\begin{equation}
d\approx\sqrt{\frac{2\rho}{\mu_0\omega}},\label{eq:skin}
\end{equation}
where $\rho$ is the resistivity of the conductor and $\mu_0$ is the permeability of free space.\cite{Griffiths:2017}  Equation~(\ref{eq:skin}) is valid provided that $d\gg l_e$, where $l_e=m_e v_\mathrm{F}/\left(ne^2\rho\right)$ is the conduction electron mean free path and $m_e$ is the electron mass, $v_\mathrm{F}\sim \SI{e6}{\meter/\second}$ is the Fermi velocity, and \mbox{$n\sim\SI{e29}{\meter\tothe{-3}}$} is the conduction electron number density.\cite{Kittel:1996}  For copper, $\rho\approx\SI{1.7e-8}{\ohm\meter}$ such that \mbox{$l_e\approx \SI{20}{\nano\meter}$} and \mbox{$d\approx \left(\SI{2}{\micro\meter\giga\hertz\tothe{1/2}}\right)/\sqrt{f}$} where \mbox{$f=\omega/\left(2\pi\right)$}.  These order-of-magnitude estimates show that Eq.~(\ref{eq:skin}) is expected to be valid for $f\lesssim\SI{1}{\tera\hertz}$.  

Referring again to Fig.~\ref{fig:RandG}(a), the cross-sectional area through which the current in the center conductor flows is \mbox{$A_1\approx 2\pi r_1 d$}.  A similar argument can be made for the current in the outer conductor such that the total per-unit-length resistance of the coaxial cable can be estimated as
\begin{equation}
R\approx \frac{\rho_1}{A_1}+\frac{\rho_2}{A_2}=\frac{1}{2\pi}\sqrt{\frac{\mu_0\omega}{2}}\left[\frac{\sqrt{\rho_1}}{r_1}+\frac{\sqrt{\rho_2}}{r_2}\right],\label{eq:R}
\end{equation} 
where we have allowed for the possibility that the inner and outer conductors have different resistivities given by $\rho_1$ and $\rho_2$, respectively.\cite{Bobowski:2013}  Equation~(\ref{eq:R}) can be expressed more conveniently in terms of effective parameters
\begin{equation}
R=\frac{1}{2\pi r_\mathrm{eff}}\sqrt{\frac{\mu_0\omega\rho_\mathrm{eff}}{2}},\label{eq:eff}
\end{equation} 
where $r_\mathrm{eff}^{-1}=r_1^{-1}+r_2^{-1}$ and
\begin{equation}
\sqrt{\rho_\mathrm{eff}}=\frac{r_2\sqrt{\rho_1}+r_1\sqrt{\rho_2}}{r_1+r_2}.\label{eq:peff}
\end{equation}
Equation~(\ref{eq:peff}) has the desired property that the resistivity of the small-radius conductor is more heavily weighted.  The critical insight from Eqs.~(\ref{eq:R}) and (\ref{eq:eff}) is that $R\propto\omega^{1/2}$ with a constant of proportionality that is determined by the conductor resistivities and geometrical factors.

\begin{figure*}[t]
\centering{
\begin{tabular}{cc}
(a)\includegraphics[width=0.95\columnwidth]{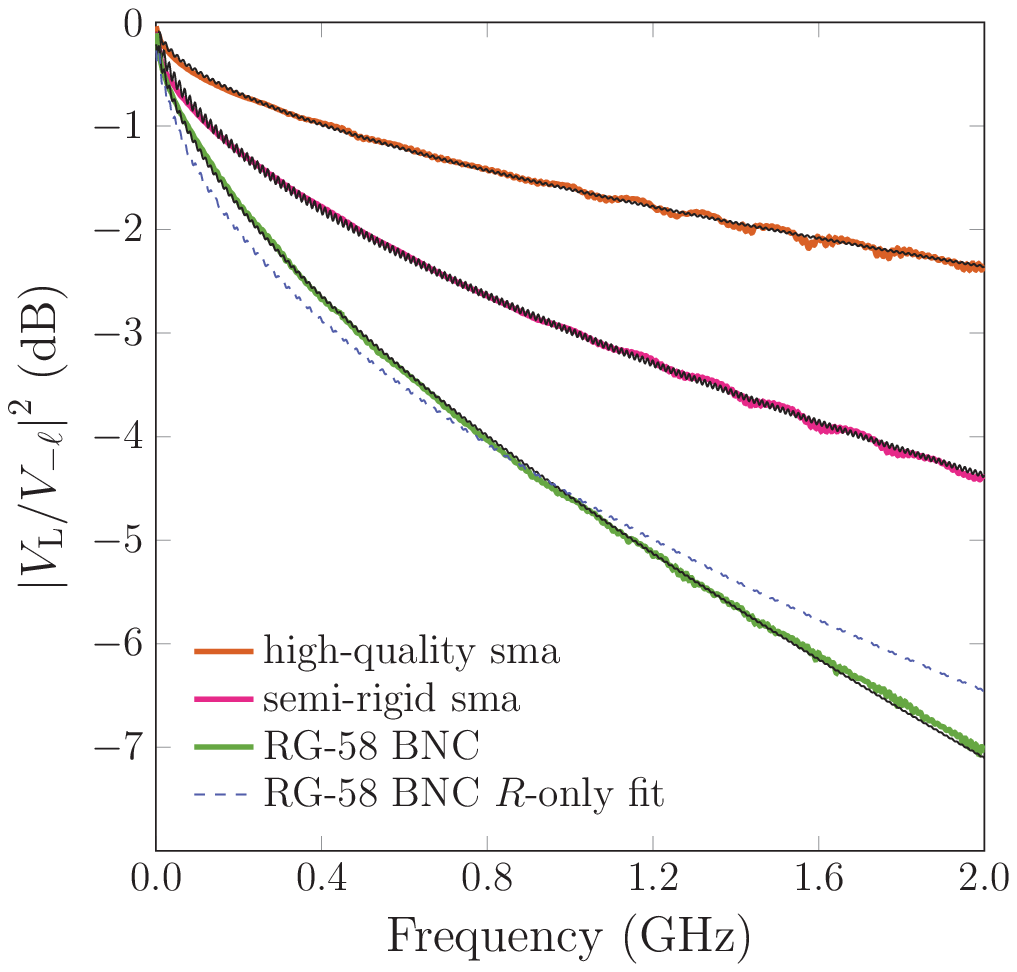} & \qquad (b)\includegraphics[width=0.92\columnwidth]{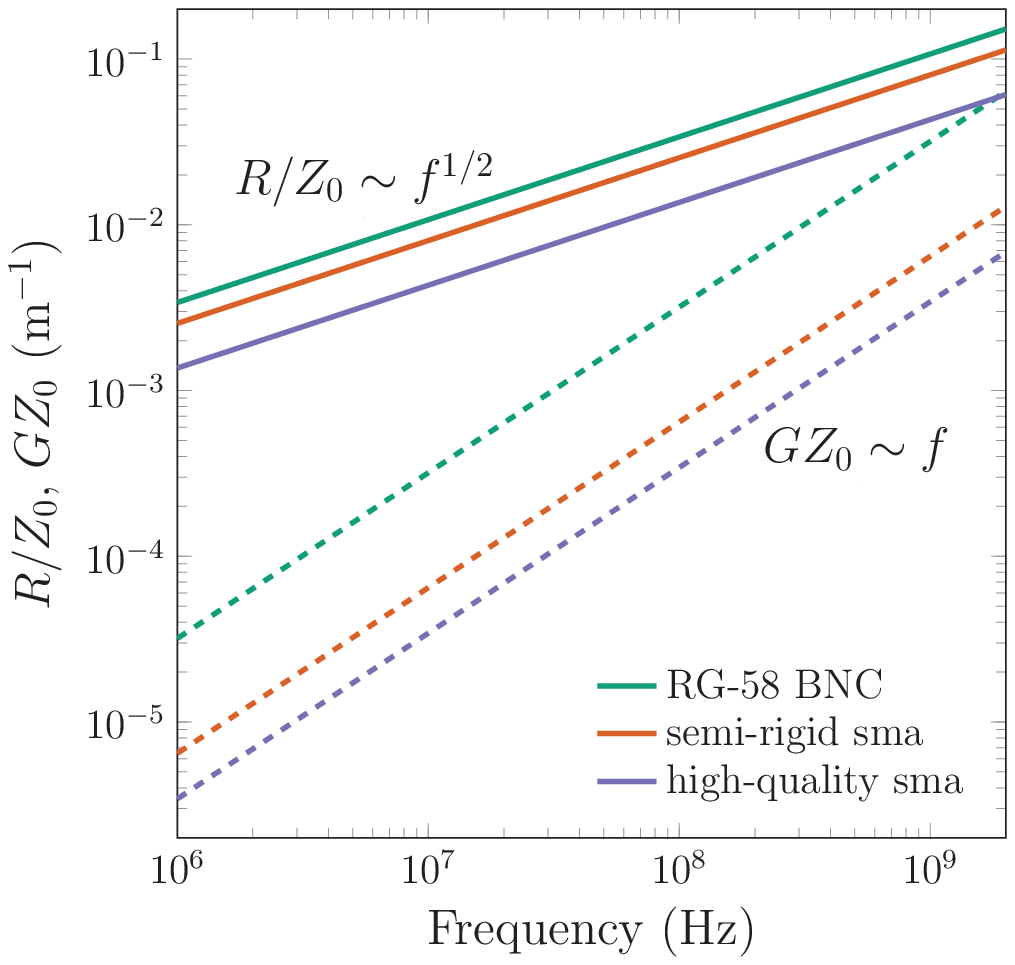}
\end{tabular}
}
\caption{\label{fig:insertion}(a) The measured insertion loss of the three coaxial transmissions lines tested.  The fits to the data (black lines), using $\left\vert V_\mathrm{L}/V_{-\ell}\right\vert^2$ calculated from Eq.~(\ref{eq:long}) and assuming $R\sim f^{1/2}$ and $G\sim f$, are excellent.  The dashed line is a fit to the RG-58 BNC cable data assuming $R/Z_0\gg GZ_0$ at all measurement frequencies.  The fit returned $a=\SI[{scientific-notation = true, separate-uncertainty = true}]{2.055(3)e-4}{\ohm\,\second\tothe{1/2}/\meter}$.  
(b) The $R/Z_0$ (solid lines) and $GZ_0$ (dashed lines) values extracted from the fits in (a) plotted as a function of frequency.}
\end{figure*}

Next, we turn our attention to $G$.  A schematic diagram of the coaxial cable cross-section is shown in Fig.~\ref{fig:RandG}(b).  The per-unit-length capacitance of a coaxial cable is given by
\begin{equation}
C=\frac{2\pi\varepsilon_\mathrm{r}\varepsilon_0}{\ln\left(r_2/r_1\right)},
\end{equation}
where $\varepsilon_\mathrm{r}$ is the relative permittivity of the dielectric material filling the space between the center and outer conductors and $\varepsilon_0$ is the permittivity of free space.  For a lossy dielectric, the relative permittivity is \mbox{$\varepsilon_\mathrm{r}=\varepsilon^\prime-j\varepsilon^{\prime\prime}$} such that the capacitive admittance becomes
\begin{equation}
Y_C=j\omega C=\frac{2\pi\omega\varepsilon_0\left(j\varepsilon^\prime+\varepsilon^{\prime\prime}\right)}{\ln\left(r_2/r_1\right)}.
\end{equation}
The real term $G=2\pi\omega\varepsilon_0\varepsilon^{\prime\prime}/\ln\left(r_2/r_1\right)$ is identified as the per-unit-length conductance.  

In general, both the real and imaginary parts of $\varepsilon_\mathrm{r}$ can have their own nontrivial frequency dependencies.  The dielectrics in our coaxial cables are polytetrafluoroethylene (PTFE, Teflon) or polythene (PE).  Attenuation in a dielectric is characterized by the loss tangent \mbox{$\tan\delta=\left(\sigma+\omega\varepsilon^{\prime\prime}\right)/\varepsilon^\prime$}.  For Teflon, the conductivity $\sigma\ll\omega\varepsilon^{\prime\prime}$ at all frequencies in the microwave range such that $\tan\delta\approx\varepsilon^{\prime\prime}/\varepsilon^\prime$.  The loss tangent and $\varepsilon^\prime$ of Teflon have been measured precisely by a variety of techniques over a wide range of frequencies.  From \SI{100}{\mega\hertz} to \SI{60}{\giga\hertz}, both $\varepsilon^\prime$ and $\tan\delta$ have been shown to be independent of frequency, with $\varepsilon^\prime\approx 2.05$ and \mbox{$2\times 10^{-4}<\tan\delta<3\times 10^{-4}$}.\cite{Givot:2006, Krupka:1996, Seo:2002, Humbert:1996, Afsar:2001}  In the analysis presented in Sec.~\ref{sec:fitting}, which includes measurements that span \SI{1}{\mega\hertz} to \SI{2}{\giga\hertz}, we assume a frequency-independent $\varepsilon^{\prime\prime}$ such that $G\propto\omega$ with the constant of proportionality determined by $\varepsilon^{\prime\prime}$ and geometrical factors.

\subsection{Power ratio measurements}\label{sec:fitting}
The simple experimental setup shown schematically in Fig.~\ref{fig:distributed}(a) was used to measure the rms power $P_\mathrm{L}$ delivered to a load impedance at the end of a transmission line.  The signal source used was a Rohde \& Schwarz SMY 02 signal generator with a \SI[number-unit-product=\text{-}]{50}{\ohm} output impedance and the load impedance was a Boonton 41-4E power sensor with a \SI[number-unit-product=\text{-}]{50}{\ohm} input impedance coupled with a Boonton Model \#42BD power meter.  The dc recorder output of the power meter was monitored using a Keysight 34401A digital multimeter.  A simple LabVIEW program was written to scan the frequency of the signal generator while writing the multimeter data to a file.  The program made repeated measurements of the power at each frequency and the average of the values was recorded.  Twenty averages were used in the measurements reported in this section.

The calibration of the Boonton power sensor is frequency dependent and there can be small variations in the power output by the signal generator when sweeping over a wide frequency range.  To remove both of these effects from the measured data, we repeated the same frequency scan and measured $P_{-\ell}$ with the power meter connected directly to the output of the signal generator.  The ratio $P_\mathrm{L}/P_{-\ell}$ is independent of both the calibration of the power sensor and small variations in the output power.  Note that this power ratio is equivalent to the scattering parameter $S_{21}$ which can measured quickly and precisely using a vector network analyzer (VNA).  The results of the measurements for the RG-58 BNC, semi-rigid UT-141, and HQ sma cables are shown in Fig.~\ref{fig:insertion}(a).  The data are shown on a decibel scale and are a measure of the insertion loss resulting from the transmission lines.  The rms power ratio $P_\mathrm{L}/P_{-\ell}$ is also equivalent to $\left\vert V_\mathrm{L}/V_{-\ell}\right\vert^2$ which can be calculated from Eq.~(\ref{eq:long}).  Table~\ref{tab:specs} shows that the measured insertion losses are in good agreement with manufacturer specifications.\cite{Pasternack:2017, MicroCoax}
\begin{table}\caption{\label{tab:specs}Measured insertion losses of the RG-58, UT-141, and HQ sma coaxial cables at various frequencies.  The manufacturer specifications, when given, are shown in brackets.}
\begin{tabular}{ccccc}
\hline\hline\\ [-1.5ex]
~ & ~~{\bf \SI[detect-weight=true, detect-family=true]{10}{\mega\hertz}}~~ & ~~{\bf \SI[detect-weight=true, detect-family=true]{0.1}{\giga\hertz}}~~ & ~~{\bf \SI[detect-weight=true, detect-family=true]{0.5}{\giga\hertz}}~~ & ~~{\bf \SI[detect-weight=true, detect-family=true]{1}{\giga\hertz}}~~\\[0.5ex]
\hline\\ [-1.5ex]
RG-58 & 0.045 & 0.15 & 0.40 & 0.61\\
(\SI{}{\decibel/\meter}) & (0.046) & (0.16) & ~ & (0.66)\\[0.5ex]
\hline\\ [-1.5ex]
UT-141 & 0.040 & 0.11 & 0.25 & 0.37\\
(\SI{}{\decibel/\meter}) & ~ & ~ & (0.26) & (0.39)\\[0.5ex]
\hline\\ [-1.5ex]
HQ sma & 0.021 & 0.063 & 0.14 & 0.20\\
(\SI{}{\decibel/\meter}) & ~ & ~ & ~ & ~\\[0.5ex]
\hline\hline
\end{tabular}
\end{table} 

Fits to the insertion loss data, assuming $R=af^{1/2}$ and $G=bf$ are also shown in Fig.~\ref{fig:insertion}(a).  For all but the RG-58 cable, the $v_0=1/\sqrt{LC}$ and $Z_0=\sqrt{L/C}$ values obtained from the transient response analysis were used such that $a$ and $b$ were the only fit parameters.  The fits are are in good agreement with the developed theory but do not capture the $\approx\SI[number-unit-product=\text{-}]{175}{\mega\hertz}$ modulations that emerge in the measured data above \SI{1}{\giga\hertz}.  The source of these modulations has not been identified.  The values of $a$ and $b$ extracted from the fits are given in Table~\ref{tab:fits}.
\begin{table}\caption{\label{tab:fits}Parameters extracted from fits to the insertion loss data shown in Fig.~\ref{fig:insertion}(a).  For the RG-58 cable, an additional $Z_0$ fit parameter was used.  For the UT-141 cable, estimates of $\rho_\mathrm{eff}$ and $\tan\delta=\varepsilon^{\prime\prime}/\varepsilon^\prime$ were calculated from $a$ and $b$, respectively.}
\begin{tabular}{llll}
\hline\hline\\ [-1.5ex]
~ & {\bf RG-58}\quad~ & {\bf UT-141}\quad~ & {\bf HQ sma}\\[0.5ex]
\hline\\ [-1.5ex]
$a$ (\SI{e-4}{\ohm\,\second\tothe{1/2}/\meter}) & $1.688\pm 0.003$ & $1.252\pm 0.002$ & $0.680\pm 0.001$\\
$b$ (\SI{e-13}{\second/\ohm/\meter}) & $6.42\pm 0.03$ & $1.31\pm 0.03$ & $0.69\pm 0.01$\\
$Z_0$ (\SI{}{\ohm}) & $49.71\pm 0.02$ & ~ & ~\\
$\rho_\mathrm{eff}$ (\SI{e-8}{\ohm\meter}) & ~ & $1.94$ & ~\\
$\varepsilon^{\prime\prime}/\varepsilon^\prime$ (\SI{e-4}{}) & ~ & $2.02$ & ~\\[0.5ex]
\hline\hline
\end{tabular}
\end{table} 
For the semi-rigid UT-141, for which $r_1$ and $r_2$ are precisely known, we estimated the values of $\rho_\mathrm{eff}$ and $\tan\delta=\varepsilon^{\prime\prime}/\varepsilon^\prime$ using
\begin{align}
\rho_\mathrm{eff} &=\frac{\left(a\, r_\mathrm{eff}\right)^2}{\mu_0/\left(4\pi\right)}\\
\tan\delta &=\frac{b\ln\left(r_2/r_1\right)}{4\pi^2\varepsilon_0\varepsilon^\prime}.
\end{align}
The results of these estimates using \mbox{$r_1=\SI{0.460}{\milli\meter}$}, \mbox{$r_2=\SI{1.493}{\milli\meter}$}, and \mbox{$r_\mathrm{eff}=\SI{0.352}{\milli\meter}$} are given in Table~\ref{tab:fits}.  

The center conductor of the UT-141 cable is silver-plated copperweld (SPCW).  Copperweld is a wire in which a copper cladding is bonded to a steel core.  The conductivity of the cladding can be anywhere from \SI{30}{} to \SI{70}{\percent} IACS (International Annealed Copper Standard, $\SI{58.2e6}{\ohm\tothe{-1}\meter\tothe{-1}}$).  The silver plating will reduce the overall effective resistivity which, depending on the thickness of the plating, could be frequency dependent due to the changes in the EM skin depth.  Given these considerations, the extracted estimate of $\rho_\mathrm{eff}$, being about \SI{15}{\percent} greater than that of copper, is very reasonable.

The UT-141 coaxial cable has a Teflon dielectric.  The extracted value of $\tan\delta$ falls directly within the range of values reported in the literature.\cite{Givot:2006, Krupka:1996, Seo:2002, Humbert:1996, Afsar:2001}  This result is remarkable because typically precision resonator techniques are required to accurately measure dissipation in low-loss dielectrics.     

\begin{figure}[t]
\centering{
\includegraphics[width=\columnwidth]{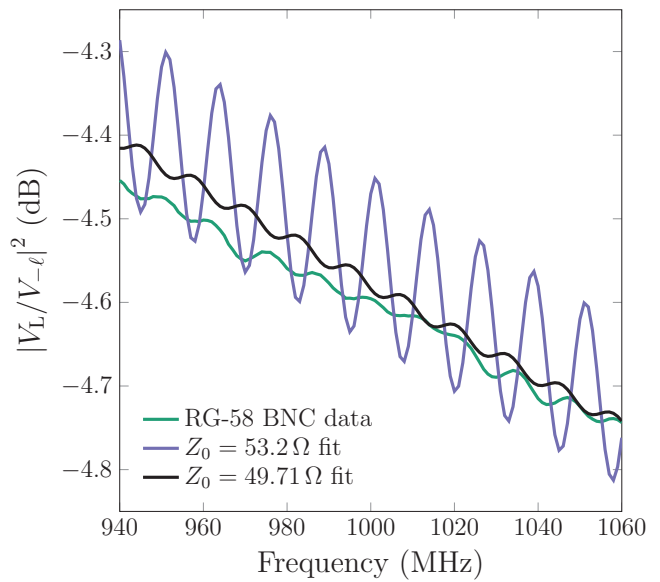}
}
\caption{\label{fig:oscillations}Insertion loss frequency oscillations.  The green line shows the same RG-58 BNC data displayed in Fig.~\ref{fig:insertion}(a), but over a narrow frequency range.  The purple line is a fit to the data using a fixed value of $Z_0=\SI{53.2}{\ohm}$ obtained from the transient analysis.  The black line is a fit in which $Z_0$ was included as a fit parameter.}  
\end{figure}

When fitting the insertion loss data of the RG-58 cable, the characteristic impedance of \SI{53.2}{\ohm} obtained from the transient analysis resulted in relatively large frequency oscillations in the fit function.  A detailed view of the data and the oscillations are shown in Fig.~\ref{fig:oscillations}. 
We speculate that dissipation in the RG-58 cable is non-negligible such that the lossless transmission line approximation used in the transient analysis breaks down.  If we assume that $V_\mathrm{g}$ measured in the transient analysis is reduced by \SI{0.2}{\percent} each time the signal travels the length of the cable and back, then we find that the corrected data results in a characteristic impedance of \SI{50}{\ohm}.  The black lines in Figs.~\ref{fig:insertion}(a) and \ref{fig:oscillations} are fits to the RG-58 data in which $Z_0$ was included as an additional fit parameter.  The best-fit value returned was $Z_0=\SI{49.71}{\ohm}$ which results in oscillation amplitudes that are much closer to those observed in the measured data.  We note that it would also be possible to deduce the characteristic impedance of the line using Eq.~(\ref{eq:G}) and a separate measurement of the reflection coefficient $\Gamma$ as a function of the load impedance $Z_\mathrm{L}$.

If $Z_0$ is included as a parameter in the fits to the semi-rigid and HQ sma data, a fit value that is in experimental agreement with results from the transient analysis are returned.  Finally, we note that the average period of the observed oscillations highlighted in Fig.~\ref{fig:oscillations} agrees very well with the expected value of $v_0/\left(2\ell\right)=\SI{12.5}{\mega\hertz}$ based on the transient analysis.

Figure~\ref{fig:insertion}(b) shows plots of the frequency dependencies of $R$ and $G$ for all of the transmission lines measured using the values of $a$ and $b$ extracted from the fits (see Table~\ref{tab:fits}).  At the lowest frequencies, $R/Z_0$ is at least two orders of magnitude greater that $GZ_0$ such that $\alpha_+\approx\alpha_-\approx R/Z_0$.  However, because $G$ increases more quickly with frequency than $R$, its contribution to the overall insertion loss becomes more important as frequency is increased.  In the case of the RG-58 cable, dielectric losses match conductor losses, and then surpass them, by \SI{2}{\giga\hertz}.  This crossover occurs by \SI{25}{\giga\hertz} for the UT-141 and HQ sma cables.  The dashed line in Fig.~\ref{fig:insertion}(a) shows a fit to the insertion loss of the RG-58 cable assuming that $G$ is negligible.  Clearly, the dielectric losses need to be included to capture the full frequency dependence of the measured insertion loss.

Finally, we note that at high frequencies where dielectric losses dominate or in situations requiring very low attenuation, air-dielectric coaxial cables are available.  In one example, a helical polyethylene spacer is used to keep the inner conductor concentric with the outer conductor, both of which are made from corrugated copper.  The insertion loss of the HELIFLEX air-dielectric coaxial cable is specified to be \SI{0.021}{\decibel/\meter} at \SI{1}{\giga\hertz} which is an order of magnitude less than the value found for the ``high-quality'' sma cable characterized in this work.\cite{HELIFLEX:2007}  Waveguides can also be used in low-loss applications.  For example, Mega Industries offers WR950 waveguide that can be used between \num{0.75} and \SI{1.12}{\giga\hertz} and has an insertion loss less than \SI{0.005}{\decibel/\meter}.\cite{Mega}

\section{Summary}\label{sec:summary}
We have described two experiments that, together, can be used to fully characterize the properties of transmission lines.  Both experiments are simple to set up and make use of equipment that is either commonly available in undergraduate labs or relatively inexpensive to acquire.

First, the transient response to a voltage step was used determine the transmission line signal propagation speed and characteristic impedance or, equivalently, the per-unit-length capacitance and inductance.  In this well-known experiment, the data analysis assumes that the transmission lines are lossless.  We found that this approximation worked well for the relatively low-loss semi-rigid UT-141 and high-quality sma coaxial cables.  However, we found evidence suggesting that losses in the RG-58 coaxial cable were causing the characteristic impedance to be systematically overestimated.

Our main objective was to use insertion loss measurements to determine the per-unit-length resistance and conductance of the same coaxial cables used in the first experiment.  With one end of the transmission line driven by a sinusoidal voltage source, and assuming small but non-negligible losses, an expression for the power delivered to a load termination $Z_\mathrm{L}$ was derived.  Based on the geometry of the coaxial cables, the frequency dependence of the conductor losses was assumed to be determined by the EM skin depth such that $R\propto f^{1/2}$.  An analysis of the dielectric losses, due to $\varepsilon^{\prime\prime}$, was used to deduce that $G\propto f$.   

The insertion loss measurements, where possible, were compared to manufacturer specifications and found to be in good agreement.  Fits to the data using the theoretical model developed were excellent.   It was shown that both the $R$ and $G$ contributions were required to capture the full frequency dependence of the measured insertion losses.  The parameters extracted from the semi-rigid UT-141 fit were used to make reasonable estimates of the cable's effective resistivity and the loss tangent of the Teflon dielectric.  

These measurements also serve to highlight the importance of the non-ideal characteristics of transmission lines, which are often not emphasized in theoretical treatments at the undergraduate level.  For example, despite impedance matching at the source and load, at \SI{2}{\giga\hertz} only \SI{20}{\percent} of the incident power is delivered to the $Z_\mathrm{L}$ termination at the end of a \SI[number-unit-product=\text{-}]{7.61}{\meter} long RG-58 coaxial cable.  For the highest-quality cable characterized in our measurements ($\ell=\SI{8.04}{\meter}$), although the power transfer efficiency increased, at \SI{58}{\percent} efficiency, substantial attenuation was still observed.

\appendix*

\section{Parts and Suppliers}
This appendix provides a list of the equipment, with possible vendors and prices, required to reproduce all parts of the transient response and insertion loss experiments described in this paper.

{\it RG-58 BNC Coaxial Cable} --  A 25-foot length (\SI{7.6}{\meter}) of RG-58 coaxial cable fitted with BNC connectors can be purchased for \SI{24}[\$]{} from Digi-Key Electronics \mbox{(\url{https://www.digikey.com/})}.

{\it Semi-Rigid Coaxial Cable} --  A 25-foot length (\SI{7.6}{\meter}) of RG-402 (UT-141) semi-rigid coaxial cable fitted with sma connectors can be purchased for \SI{234}[\$]{} from Pasternack Enterprises \mbox{(\url{https://www.pasternack.com})}.

{\it Pusle Generator} --  The HP 8011A pulse generator used for the transient response experiment is obsolete.  A possible substitute is the Rigol DG812 \SI{10}{\mega\hertz} signal generator which can be purchased for \SI{299}[\$]{} from Digi-Key Electronics.  Note that it is also possible to create a suitable voltage step using only a PP3 \SI[number-unit-product=\text{-}]{9}{\volt} battery and jumper wires.

{\it Oscilloscope} -- We used a Tektronix TBS 1104 oscilloscope to record the voltage transients.  The two-channel version of this oscilloscope (TBS 1102) can be purchased for \SI{1190}[\$]{} from Newark \mbox{(\url{https://www.newark.com})}.  Many college and university physics departments will already have a suitable oscilloscope available.

{\it Signal Generator} -- The signal for our insertion loss measurements was provided by a Rohde \& Schwarz SMY02 \SI{9}{\kilo\hertz} to \SI{2080}{\mega\hertz} signal generator.  A suitable substitute would be the Rigol DSG815 \SI{9}{\kilo\hertz} to \SI{1.5}{\giga\hertz} signal generator which sells for \SI{2099}[\$]{} \mbox{(\url{https://www.rigolna.com})}.

{\it Power Meter/Sensor} -- Our power measurements were made a using Boonton 42BD power meter with Boonton 41-4E power sensors.  The Mini-Circuits \mbox{ZX47-50-S+} power detector can be used from \SI{10}{\mega\hertz} to \SI{8}{\giga\hertz} to measure powers from $-45$ to $\SI[retain-explicit-plus]{+15}{dBm}$ and costs \SI{90}[\$]{} \mbox{(\url{https://www.minicircuits.com/})}.

\section*{Acknowledgments}
We thank the referees for their thorough review of the manuscript and the insightful suggestions for improvements.

\end{document}